\documentclass[11pt]{article}
\usepackage{amssymb,amsfonts,amsmath,latexsym,epsf,tikz,url}
\usepackage{graphicx}
\usepackage{float}
\usepackage{fancyvrb}

\usepackage{algorithm}
\usepackage{algorithmic}

\newtheorem{theorem}{Theorem}[section]

\newtheorem{conjecture}[theorem]{Conjecture}

\newtheorem{example}[theorem]{Example}

\newtheorem{problem}[theorem]{Problem}

\newcommand{\proof}{\noindent{\bf Proof.\ }}
\newcommand{\qed}{\hfill $\square$\medskip}

\begin{document}

\title{Graphs and codes produced by a new method for dividing a natural number by two}

\author{
M. Zeynali Azim$^a$, M.A. Jabraeil Jamali$^{a,}$\footnote{Corresponding author}, B. Anari$^a$, S. Alikhani$^b$, S. Akbarpour $^a$ }
\date{}

\maketitle

\begin{center}
	$^a$Department of Computer Engineering, Shabestar Branch, Islamic Azad University, Shabestar, Iran
	
	{\tt  m\_jamali@itrc.ac.ir }
	\medskip 
	
$^b$Department of Mathematics, Yazd University, 89195-741, Yazd, Iran\\

\end{center}


\begin{abstract}
In this paper, we introduce a new method which we call it MZ-method, for dividing a natural number $x$ by two and then  we use graph as a model to show MZ-algorithm. Applying (recursively)  $k$-times of the MZ-method for the number $x$, produces a  graph with unique structure that is denoted by $G_k(x)$. We investigate the structure of $G_k(x)$.   Also from the natural number $x$ and graph $G_k(x)$ we produce codes which are important and applicable in the information security and cryptography.  
\end{abstract}

\noindent{\bf Keywords:} graph, division, code.

\medskip

\section{Introduction}

A division algorithm is an algorithm which, given two integers $N$ and $D$, computes their quotient and/or remainder, the result of Euclidean division. Some are applied by hand, while others are employed by digital circuit designs and software. 
For a real number $x$ and an integer $n\geq 0$, let $[x]_n$ denote the (finite) decimal expansion of the greatest number that is not greater than $x$, which has exactly $n$ digits after the decimal mark. Let $d_i$ denote the last digit of $[x]_i$. It is straightforward to see that $[x]_n$ may be obtained by appending $d_n$ to the right of $[x]_{n-1}$. This way one has
\[
[x]_n = [x]_0.d_1d_2...d_{n-1}d_n,
\]
It is obvious that for an integer $x$,  $[\frac{x}{2}]_n=[\frac{x}{2}]_0.d_1$, where $d_1=0$, or $d_1=5$.  
In mathematics, division by two or halving has also been called mediation or dimidiation \cite{1}. The treatment of this as a different operation from multiplication and division by other numbers goes back to the ancient Egyptians, whose multiplication algorithm used division by two as one of its fundamental steps. Some mathematicians as late as the sixteenth century continued to view halving as a separate operation, and it often continues to be treated separately in modern computer programming. Performing this operation is simple in decimal arithmetic, in the binary numeral system used in computer programming, and in other even-numbered bases \cite{3,4}.

In binary floating-point arithmetic, division by two can be performed by decreasing the exponent by one (as long as the result is not a subnormal number). Many programming languages provide functions that can be used to divide a floating point number by a power of two. For example, the Java programming language provides the method java.lang.Math.scalb for scaling by a power of two, and the C programming language provides the function ldexp for the same purpose \cite{wiki}. 
There are some algorithms for division by two. For instance see \cite{wiki}. 

\medskip

In this paper, we introduce a new method which we call it MZ-method, for dividing a natural number $x$ by two. With our knowledge this method is not in any literature. Then  we use graph as a model to show MZ-algorithm. Applying  $k$-times of the MZ-method for the number $x$, creates  a  graph with unique structure that is denoted by $G_k(x)$. We investigate the structure of $G_k(x)$ in the next section. Also from the natural number $x$ and graph $G_k(x)$, in Section 3,  we produce codes which are important and applicable in the information security and cryptography.

\section{A new algorithm for division by two}
In this section, we first present a new algorithm which we call it MZ-algorithm for division by two. 

\bigskip
STEP 1: Consider an input natural number $x$ and write $x$ as $\overline{a_{n-1}a_{n-2}...a_1a_0}$.
 \bigskip
 
STEP 2:  Write $\frac{a_j}{2}$ as $b_j t_j$ ($0\leq j\leq n-1$), where $b_j=\lfloor \frac{a_j}{2}\rfloor$ and 
 $t_j=\big(\frac{a_j}{2}-\lfloor \frac{a_j}{2}\rfloor\big)\times 10$.
 
 \bigskip
STEP 3:  $\frac{x}{2}=\overline{c_{n}c_{n-1}...c_1.c_0}$, where 
 $c_n=\lfloor\frac{a_{n-1}}{2}\rfloor$,  $c_0=t_0$ and $c_j=b_{j-1}+t_{j}$ for $1\leq j\leq n-1$. 
 
 \begin{figure}[H]
 	\begin{center} 
 		\includegraphics[width=9cm, height=7.20cm]{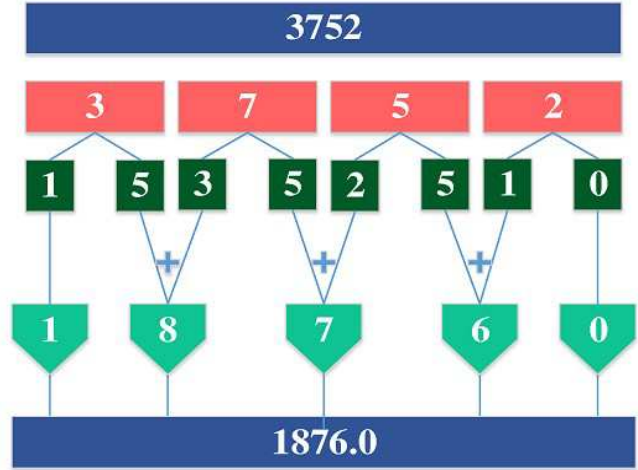}
 	\end{center} 
 	\caption{\label{fig0} Algorithm for the devision number $375$ by two.  }
 \end{figure}

 \bigskip
 
 To prove this algorithm, we state and prove the following theorem: 
 \begin{theorem} 
 	Let $x=\overline{a_{n-1}a_{n-2}...a_1a_0}$ be a natural number. Then $$\frac{x}{2}=\overline{c_{n}c_{n-1}...c_1.c_0},$$
 	 where 
 	$c_n=\lfloor\frac{a_{n-1}}{2}\rfloor$,  $c_0=\big(\frac{a_0}{2}-\lfloor \frac{a_0}{2}\rfloor\big)\times 10$ and $c_j=\lfloor \frac{a_{j-1}}{2}\rfloor+
 	\big(\frac{a_{j}}{2}-\lfloor \frac{a_{j}}{2}\rfloor\big)\times 10$. 
 	\end{theorem} 

\proof
We know that the number $x$ is equal to $a_0+a_1\times 10+a_2\times 10^2+...+a_{n-1}\times 10^{n-1}$ and so 
\begin{eqnarray}\label{1}
\frac{x}{2}=\frac{a_0}{2}+a_1\times 5+a_2\times 50+...+a_{n-1}\times 5\times 10^{n-2}.
\end{eqnarray}
We shall show that $\frac{x}{2}=\overline{c_{n}c_{n-1}...c_1.c_0}$. We write this number as
$$c_n\times 10^{n-1}+c_{n-1}\times 10^{n-2}+...+c_1+\frac{c_0}{10}.$$ 
By substituting $c_j$, we have:

$\lfloor\frac{a_{n-1}}{2}\rfloor\times 10^{n-1}+\Big(\lfloor \frac{a_{n-2}}{2}\rfloor+
\big(\frac{a_{n-1}}{2}-\lfloor \frac{a_{n-1}}{2}\rfloor\big)\times 10\Big)\times 10^{n-2}+...\\$\begin{eqnarray}\label{2} 
+\lfloor \frac{a_{0}}{2}\rfloor+\frac{a_{1}}{2}-\lfloor \frac{a_{1}}{2}\rfloor \times 10.
\end{eqnarray}
Obviously the Equation (2) is equal to Equation (1) and so we have the result.\qed

\medskip 
As an example, Figure \ref{fig0} illustrates the MZ-algorithm for division number $375$ by two.

\section{Graphs and codes from MZ-algorithm }

In this section, we use MZ-algorithm recursively and produce graphs and codes. 

\begin{figure}
 	\begin{center} 
 		\includegraphics[width=11cm, height=4.25cm]{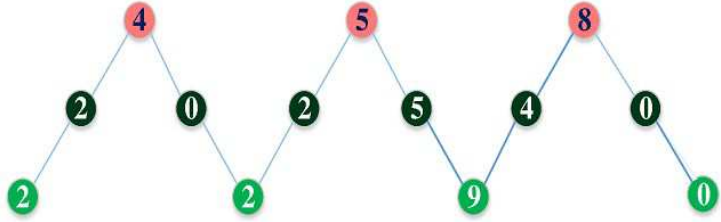}
 	\end{center} 
 	 \caption{\label{fig00} The graph  $G_1(458)$.  }
 \end{figure}

 A graph is a pair $G=(V,E)$, where $V$ is the vertex set and $E$ is an edge set. Let to use graph as a model for MZ-algorithm. Every digit in  a number denoted by a vertex and edges of graph draw based on MZ-algorithm. We have shown the division of the number 458 by two in Figure \ref{fig00}. This graph (Figure \ref{fig00}) which we call it division graph by two (DGBT) is a path of order $13$, i.e., $P_{13}$.    
 
\medskip
Applying (recursively)  $k$-times of the MZ-method for the number $x$, produces a  graph which we denote it by $G_k(x)$.  
The following theorem is easy to obtain: 

\begin{theorem}\label{thm1} 
Let $x$ be a natural number with $n$ digit.	The division graph $D_1(x)$  is a path graph $P_{4n+1}$. 
\end{theorem}

\begin{figure}
	\vspace{-.05cm}
	\begin{center} 
		\includegraphics[width=11cm, height=4cm]{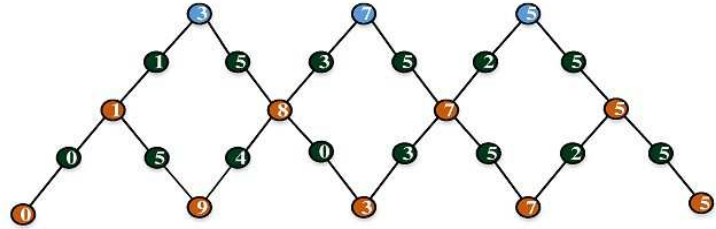}
		\caption{\label{fig2} The graph  $G_2(375)$.}
	\end{center} 
\end{figure}

\begin{figure}
	\vspace{-.05cm}
	\begin{center} 
		\includegraphics[width=12cm, height=5cm]{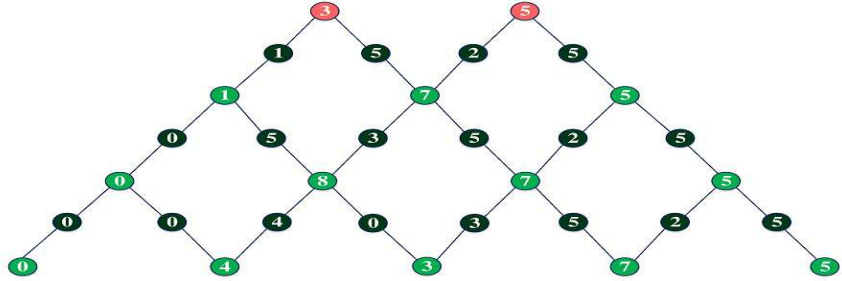}
	\end{center} 
	\caption{\label{fig3} The graph $G_3(35)$.  }
\end{figure}

It is easy to see that $G_k(n)$ is not tree for $k>1$, since the  graph has cycle. See the graphs 
$G_2(375)$ and  $G_3(35)$ in Figures \ref{fig2} and \ref{fig3}.
Since DGBT is an infinite graph, let to show some of DGBT in bitmap model. In bitmap model each numbers in ${0,1,2,...,9}$ represented by unique color. See Figures \ref{fig4} and \ref{fig5}.

\begin{figure}[H]
	
	\begin{center} \label{fig4}
		\includegraphics[width=11.2cm, height=6cm]{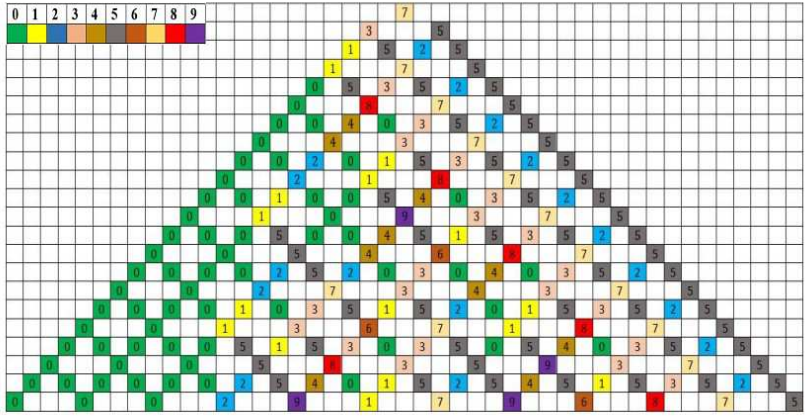}
	\end{center} 
	\caption{\label{fig4} Bitmap model for the graph  $G_{11}(7)$ }
\end{figure}

\begin{theorem} Let $x$ be a $d$-digit natural number. 
	\begin{enumerate}
		\item[(i)] The order of graph $G_k(x)$ is $(3k+1)d+\frac{k}{2}(3k-1)$.
		\item[(ii)] The number of cycle $C_8$ in the graph $G_k(x)$ is $\frac{k-1}{2}(2d+k-2)$.
		\item[(iii)]  For every natural numbers $k,n$, the graph $G_k(n)$ has exactly two end vertex (leaf).
	\end{enumerate}
\end{theorem} 
\proof
\begin{enumerate}
	\item[(i)] By construction of graph $G_k(x)$, the number of vertices is equal to
	$$d+2d+(d+1)+2(d+1)+(d+2)+2(d+2)+...+2(d+k-1)+d+k,$$
	which is equal to $(3k+1)d+\frac{k}{2}(3k-1)$. 
	
	\item[(ii)]  From construction of graph $G_k(x)$ observe that the number of cycle $C_8$ is 
	$d+(d+1)+(d+2)+...+(d+k-2),$  which is equal to $\frac{k-1}{2}(2d+k-2)$.
	
	\item[(iii)]  It is straightforward.\qed
\end{enumerate}

\begin{figure}[H]
	
	\begin{center} 
		\includegraphics[width=9.1cm, height=10cm]{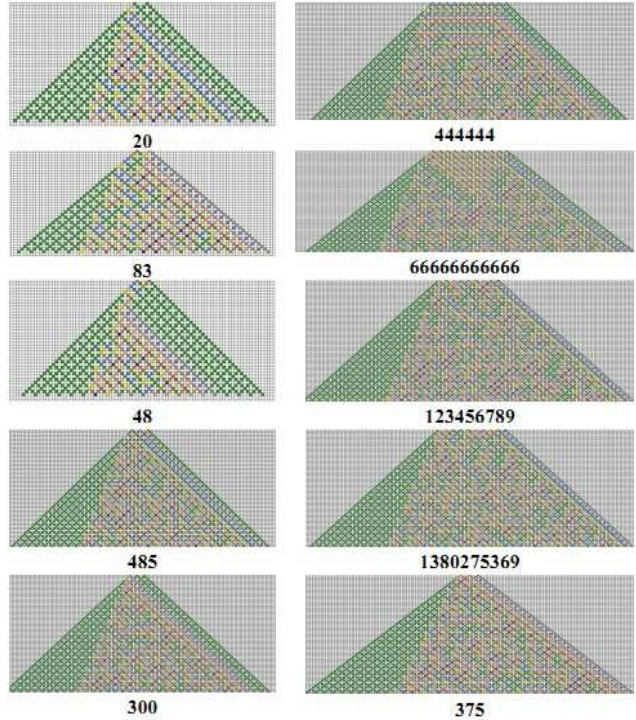}
	\end{center} 
	\caption{\label{fig5} Bitmap model for some DGBT. }
\end{figure}

\begin{figure}[h]
		\hspace{4cm}
		\begin{center} 
			\includegraphics[width=11cm, height=8.5cm]{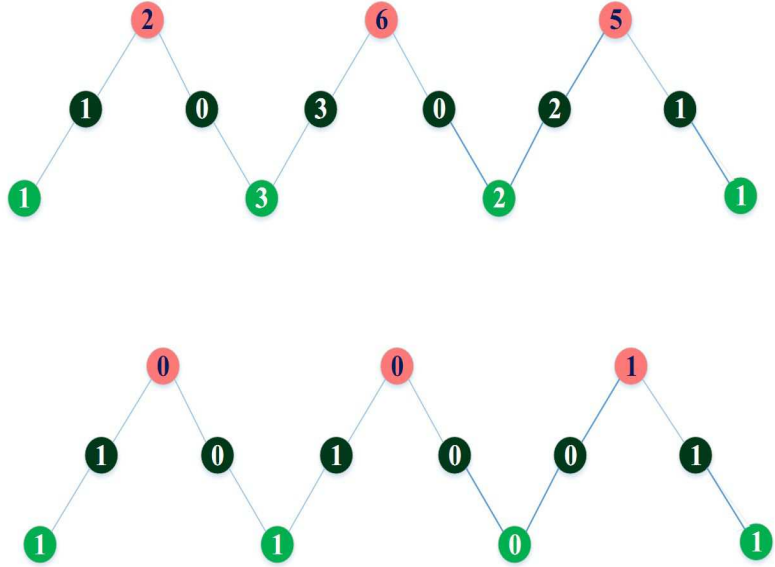}
		\end{center} 
		\caption{ \label{fig111} The $G_1(265)$ and its binary graph $BG_1(265)$.  }
	\end{figure}

	Random numbers are useful for a different of purposes, such as generating data encryption keys, simulating and modeling complex phenomena and for selecting random samples from larger data sets. They have also been used aesthetically, for example in literature and music, and are of course ever popular for games and gambling.  Our assumption has been that random numbers cannot be computed; because computers operate in deterministic way, they cannot produce random numbers. Instead, random numbers are best obtained using physical (true) random number generators \cite{random}.   MZ-algorithm produce numbers that we think can be consider as true random generator.   
	Also  we can construct  binary codes from MZ-algorithm and its associated graph. We replace labeled  numbers to vertices of  graph $G_k(x)$ by $1$ (for odd numbers) and by $0$ (for even numbers). Let to call this graph $BG_k(x)$. See Figure \ref{fig111}.  

\begin{figure}
	\begin{center} 
		\vspace{-.1cm}
		\includegraphics[width=11.5cm, height=8.4cm]{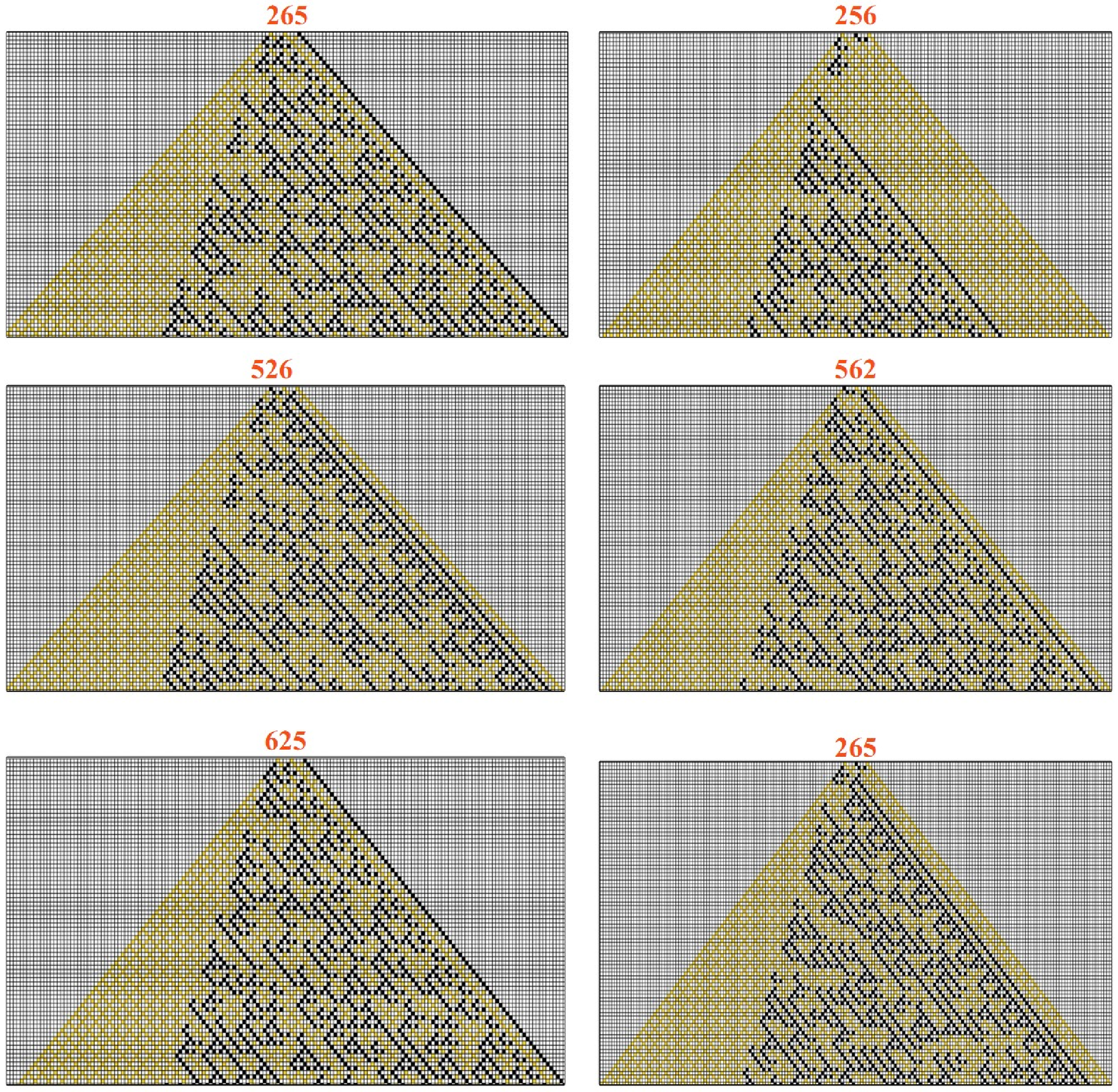}
		
			\includegraphics[width=11.5cm, height=8.4cm]{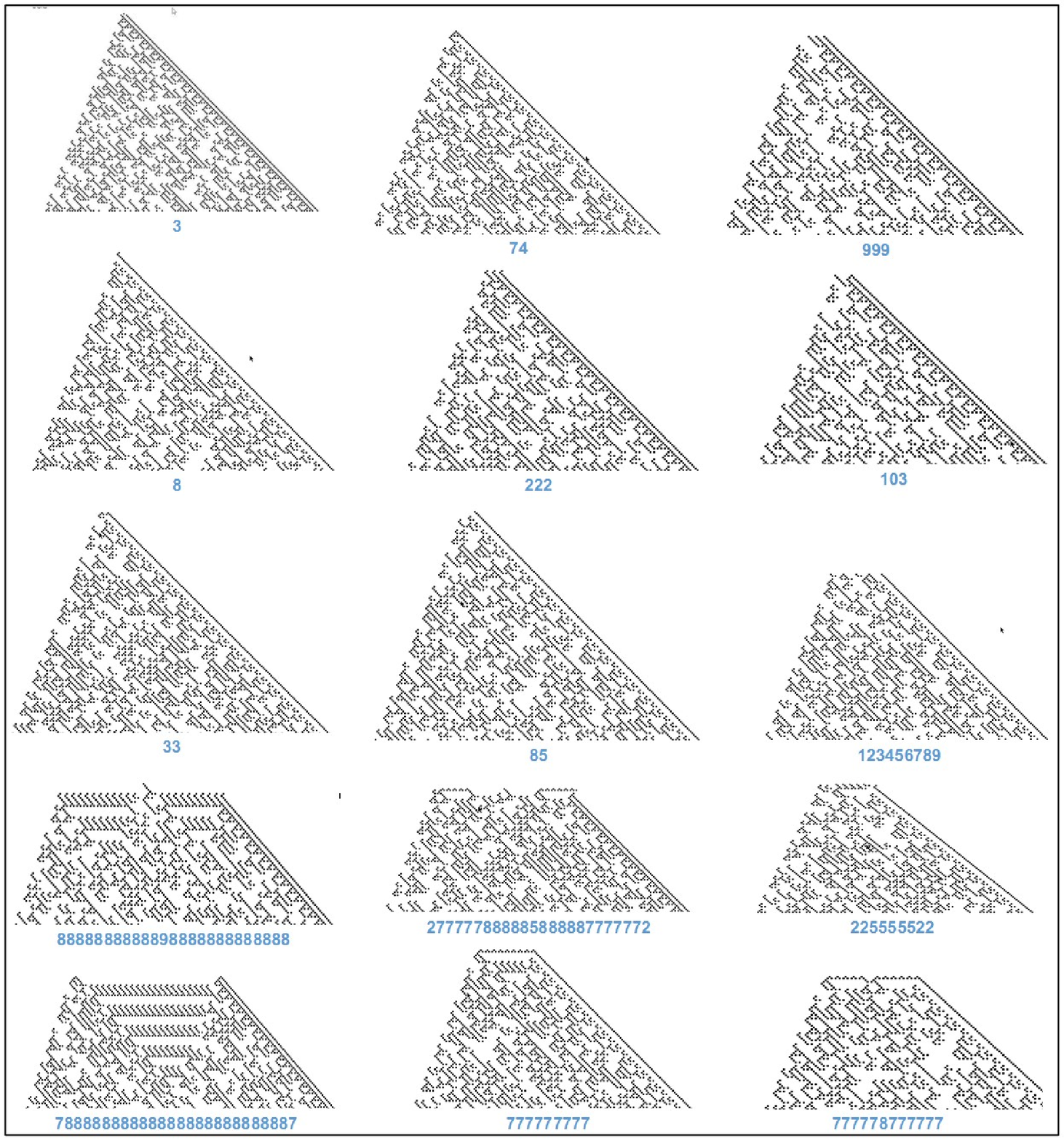}
		\caption{\label{fig21} Binary graph Visualization }
	\end{center} 
\end{figure}

As we saw in previous section, DGBT is an infinite graph so it related binary graph (BG) is also infinite graph. We have drawn some  BG in bitmap format, see Figure \ref{fig21}.  After simulating and drawing the bitmap model of BG, observe that  the MZ-algorithm BG is similar to a cellular automaton and can be defined as its  role. Also, due to the lack of specific pattern, we think that BG can be used for cryptography and generating random numbers.
We know that an $(n,M,d)$-code $C$  over $F$ is a $M$-subset of codewords  of length $n$ which $d$ is the  minimum distance in $C$. A good $(n,M,d)$ code has small $n$ and large $M$ and $d$ (see \cite{book}).
It is obvious that every natural number $x$ with $n$ digits is a codeword of length $n$ such that its weight is equal to the number of odd digits in $x$. Now by MZ-algorithm  the number $x$ is a binary codeword and using $G_k(x)$ can produce codeword of lengths $2(n+k-1)$ and   $n+k$, where $k\in \mathbb{N}$.  So from a $(n,M,d)$-code $C$ we construct 
$(n+k,M,d)$-code $C_1$ and $(2(n+k-1),M,d)$-code $C_2$.

	\begin{example}
	Consider number $x$ with  three digit $2,5,6$. We know that there are six distinct numbers with these three digits. In other words we have the following code:
	\[
	C=\{256,265,526,562,625,652\}.\]
	We can write this code as binary code as $C=\{010,001,100,001,101\}$.  Let $C_k(\{2,5,6\})$ denote the code of length $k$ which has produced by MZ-algorithm and their corresponding graphs. It is not hard to see that 
	
	$C_8(\{2,5,6\})=\{00000000,00000101,01111101,01110010,01111101,00011100\}$. 
	
	$C_9(\{2,5,6\})=\{000000000,000100001,001101001,000101010,001101001,010101100\}$
	
	$C_{10}(\{2,5,6\})=\{0000000000,0000101101,0000000101,0001100010,001101001,\\0010111100\}$
	
	Note that $C_8, C_9$ and $C_{10}$ is  $(8,6,2)$-code, $(9,6,2)$-code and $(10,6,2)$-code, respectively. 
	\end{example}

	 It is interesting that the MZ-method and corresponding graph $G_k(x)$  gives unique codewords for every $k\geq 3$. See Figure \ref{fig21}. However, until now  attempts to prove this property failed. 
	 Let to state the following conjecture:
	 
	 \begin{conjecture}
	 The MZ-algorithm and $BG_k(x)$ produce distinct binary codewords for every $k\geq 3$.	 \end{conjecture}

	\medskip
	Here we use MZ-algorithm and $BG_k$ to construct $(n,M,d)$-codes which their minimum distance is almost  $\frac{n}{2}\pm i$, where $i$ is an integer. 
	
	\medskip
	
	STEP 1: Input $t!$  natural numbers such as $x$ with $t$ distinct digit. 
	\medskip

	STEP 2: Apply MZ-algorithm for each $x$, and its result, recursively.   
	\medskip

	STEP 3: In the $k$-th step and by replacing even digit by $0$ and odd digit by $1$ we have a binary codeword $C_k(x)$ of length $t+k$.
	\medskip

	STEP 4: Delete the digit(s) $0$ before the first digit $1$ from the left of codeword $C_k(x)$ and denote this codeword as $C'_k(x)$
	\medskip

	STEP 5: For enough large $k$, the set $C'=\{C'_k(x)\}$ is a $(t+k-k_0,t!,d)$-code, where $d$ is almost $\frac{t+k-k_0}{2}$ and $k_0$ is the number of deleted $0$ in STEP 4. 

\begin{example}	Consider three numbers $2,5,6$. We Use MZ-algorithm and DBGT for each six numbers which constructed  by these numbers. For $k=10$ we have the following codewords: 
		$	C_{10} (625)= 0 0 0 0 1 0 1 1 1 1 0 0 1$ , $
	C_{10} (526)= 0 0 0 1 1 1 0 1 1 0 1 1 0$, $C_{10} (562)=0 0 0 1 0 0 0 0 0 1 0 1 0$,
	$C_{10} (652)=0 0 0 0 1 0 1 1 0 1 1 0 0$,  
$	C_{10} (265)=0 0 0 0 1 0 1 0 1 0 0 0 1$, $C_{10} (256)=0 0 0 0 1 0 0 0 0 0 0 0 0$
For more information about this code please see binary graph representation of 	$BG_{10} (265)$ for $k=10$ in Figure \ref{fig000}.

	     \begin{figure}[h]
          	\begin{center} 
	        	\includegraphics[width=11cm, height=6cm]{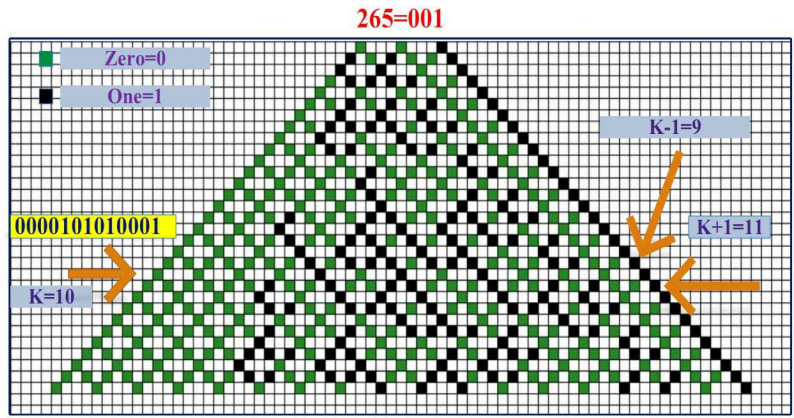}
        	\end{center} 
        	\caption{\label{fig000} Binary graph representation of number $265$ for $k=10$.}
         \end{figure}
\end{example}

\begin{figure}[H]
	\begin{center} 
		\includegraphics[width=11cm, height=6cm]{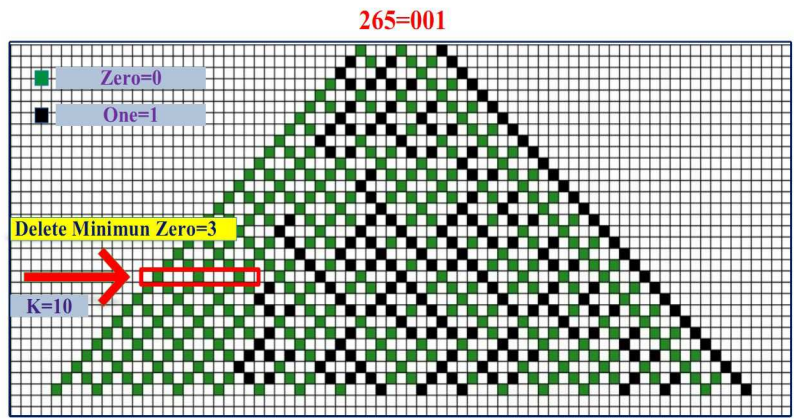}
	\end{center} 
	\caption{\label{fig10} Deleting zero from left side of $BG_{10} (265)$}
\end{figure}

By deleting three zeros from the left of these codewords, we have:

$$C'=\{0100000000, 0101010001, 1110110110, 1000001010, 0101111001, 0101101100  \}.$$
    For example, we show that deleting 3 zero from left side of $BG_{10} (265)$, see Figure \ref{fig10}.

	Note that $C'$ is a $(10,6,2)$-code. 
	We use the following procedure to generate codes:	    

             \begin{algorithm}[H]
                N[]=Get t numbers form \{0,1,...,9\} \\
				List [] =Generate $t!$ decimal number from N[] \\
				Get $k$ for generate Code \\
				While element in list[] \\
				Construct binary tree until to $L=2*k$ for any element in list[] \\
				List2[]=Get  Code from Level (L) in binary tree \\
				End of while \\
				Minimum zero= Compare all of element in list2[] and find minimum zero from left side \\
				$k_0$=minimum zero \\
				While element in list2[] \\
				While i< len(list2[]) \\
				d=find\_distance (element, list2[i+1]) \\
				d\_list[]=add(d) \\
				End of while \\
				End of while \\
				minimum\_d=find\_min(d\_list) \\
				average\_of\_code\_length=compute($(t+k-k_0)/2$) \\
				compare( minimum\_d, average\_of\_code\_length) \\
				calculate  ($t+k-k_0$, $t!$, minimum\_d)-code \\
				                \end{algorithm}

	\begin{example}
	Using this procedure we have the following results:

\begin{enumerate} 	
\item[(i)]  	$C'_{30}(\{2,5,6\})=\{010010111101100100000000,
		000000000011101000000001,\\
		001011010001101011010110,
		101001000111010101101010,
		100010001110010010001001,\\
		001000110001011110101100\}$ which is a $(24,6,10)$-code.
		\item[(ii)] $C'_{50}(\{2,5,6\})$ is  a $(38,6,16)$-code.
		\item[(iii)] $C'_{450}(\{2,5,6\})$ is a $(314,6,143)$-code.  
		\end{enumerate} 
		\end{example}

\begin{figure}[H]
		\begin{center} 
			\includegraphics[width=5cm, height=3cm]{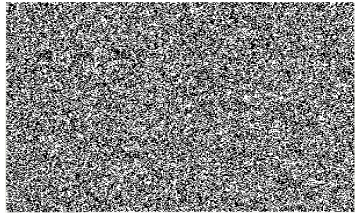}
		\end{center} 
		\caption{ \label{fig11}  Distance based bitmap of two numbers $256$, $625$, for $k=1 to k=300$. } 
	\end{figure}

\begin{figure}[H]
		\begin{center} 
			\includegraphics[width=12cm, height=14cm]{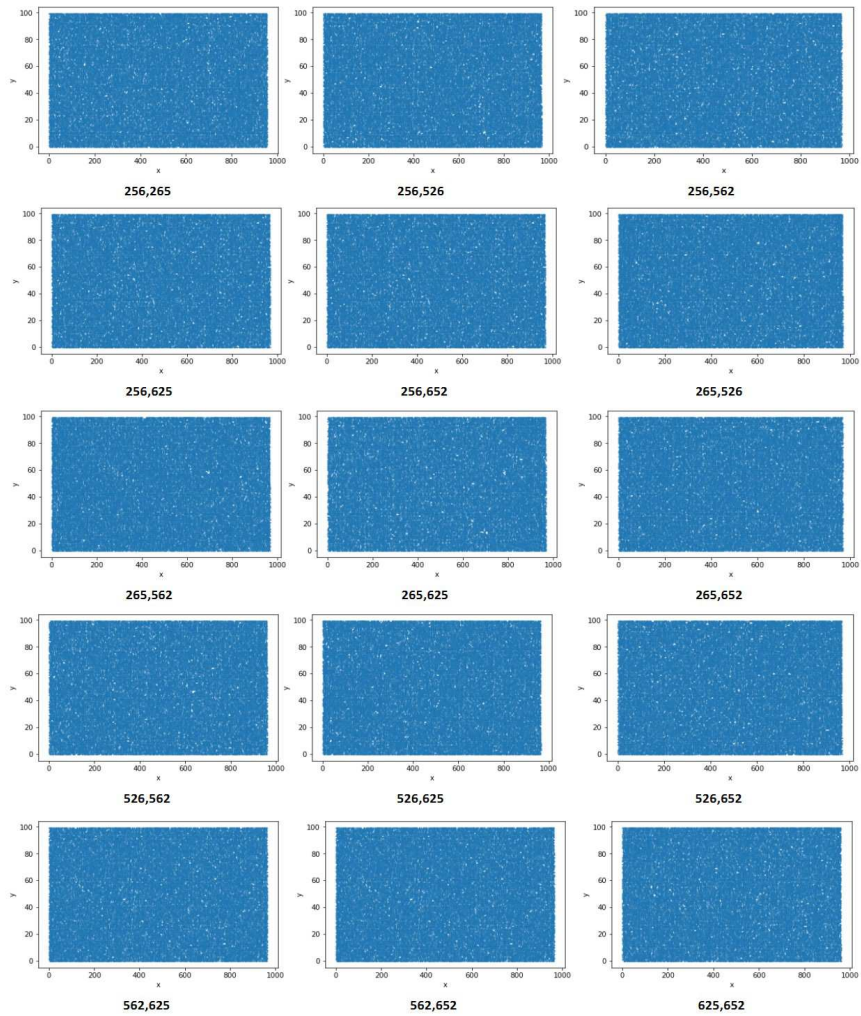}
		\end{center} 
		\caption{\label{fig12} Scatter chart for code of numbers $2,5,6$ for $k=1,...,300$. } 
	\end{figure}

	We observed that by using MZ-algorithm and BG we can produce codes of length $m$ which their minimum distance is almost $m/2$  which
	are important and applicable in the information security and cryptography. 
In the following, by the above procedure and produced codes for large enough $k$, we think that there is a characterization for number and parameters of codes for which $d$ is exactly $\frac{n}{2}$ or $\frac{n}{2}\pm i$, where $n$ is the length of code and $i$ is an integer number. Since $d$ is almost near to $\frac{n}{2}$, we were curious to draw the level to level distance based (from $k=1$ to $k=300$) bitmap model for $256$, $625$ and scatter chart for all codes from three numbers $2,5,6$. After drawing the shape, we were astonished to see that the pattern was appear to be random. So we think that  it can be used in random number generation and cryptography applications. see Figures \ref{fig11} and \ref{fig12}.

\begin{figure}[H]
	\begin{center} 
		\includegraphics[width=12cm, height=14cm]{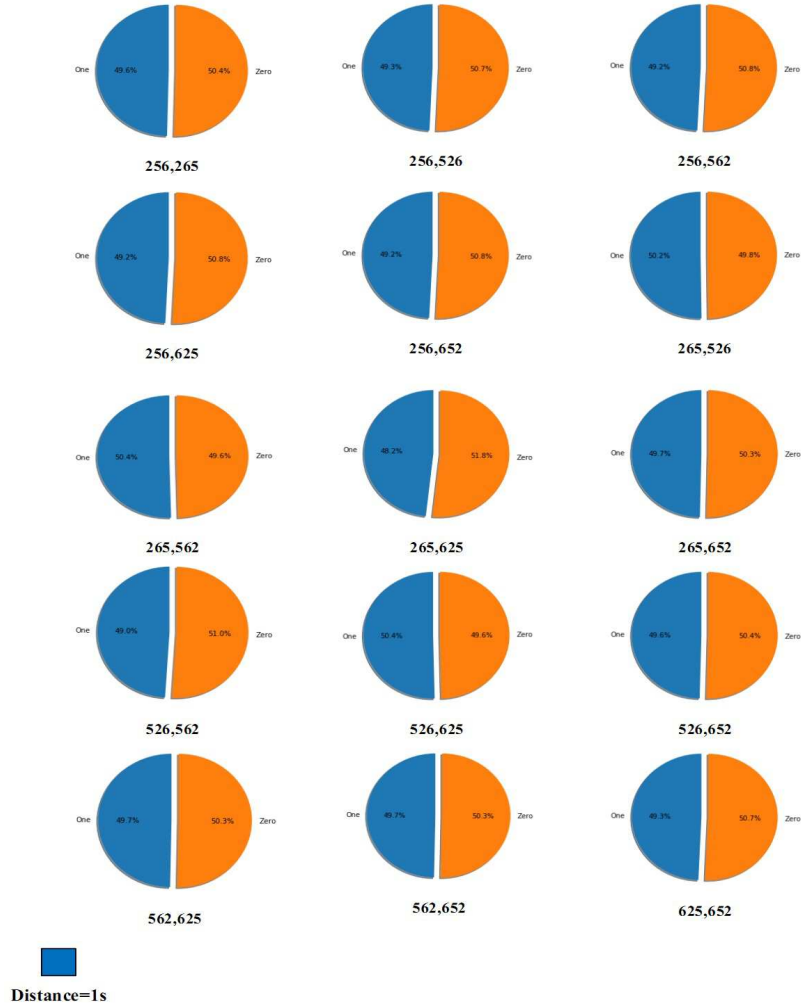}
	\end{center} 
	\caption{\label{fig1111} Pie chart for all generated codes by numbers ${2,5,6}$ based $d$. } 
\end{figure}

Also we draw pie chart for minimum distance $d$ representation in Figure \ref{fig1111}. As we can see, the number of $1$'s (or minimum distance)  is  equal or almost equal to half of length, i.e., $\frac{n}{2}$. 	However, until now all attempts to find a characterization failed.

  We close the paper with the following problem:   
	 \begin{problem}
	 	Characterize natural number $x$ and $k$ for which a produced code $(n,M,d)$ from 
	 	MZ-algorithm and DBGT, $D_k(x)$ has $d=\frac{n}{2}\pm i$, where  $i$ is an integer number. 
	 	\end{problem}



\begin{thebibliography}{99}                                                                          
		\bibitem{book} R. Hill, (1986), A first course in coding theory, Oxford University Press. 
		
		\bibitem{3} L.L. 	Jackson, (1906), The educational significance of sixteenth century arithmetic from the point of view of the present time, Contributions to education, 8, Columbia University, p. 76.
		                     	
		\bibitem{1} R.	Steele, (1922), The Earliest arithmetics in English, Early English Text Society, 118, Oxford University Press, p. 82.
		
\bibitem{random} 	M. Stip\v cevi\v c and C.K. Koc, 	True Random Number Generators, Available at: \texttt{https://www.researchgate.net/publication/299824248} 

		
	\bibitem{4} E.G.R. Waters,  (1929), A Fifteenth Century French Algorism from Liége, Isis, 12 (2): 194–236.
	
\bibitem{wiki}	See \texttt{https://en.wikipedia.org/wiki/Division\_by\_two\#cite\_note-1}, Access Jan 2020. 

	\end{thebibliography}
\end{document}